\documentclass[12pt]{article}

\usepackage{amssymb,amsmath,graphicx}

\newcommand{\tab}{\hspace{5mm}}

\begin{document}

\begin{center}
\Huge{FREQUENT ERRORS IN SPECIAL RELATIVITY}
\end{center}

\begin{center}
{\Large Diego Sa\'{a}}
\footnote{Escuela Polit\'{e}cnica Nacional. Quito --Ecuador. email: dsaa@server.epn.edu.ec}\\
\end{center}

\begin{abstract}
{Some reasons are given to suggest that the interpretation of the Lorentz' transformations as if they referred to \emph{coordinates} instead of to \emph{intervals} could be incorrect. Besides, the usual form of such transformations, by using variables that represent \emph{finite} values instead of \emph{differentials}, could be another error. Later it is shown that the Lorentz contraction factor must not have the form currently accepted for it if the Lorentz contraction factor is assumed to be equal to the quotient between time differentials.}
\end{abstract}

P.A.C.S.: \tab01.55.+b - General physics\tab03.30.+p - Special relativity

\section{THE INFINITESIMAL CHARACTER OF THE LORENTZ TRANSFORMATIONS}

\indent A seemingly overlooked error, preserved since the original works of Lorentz, 
Minkowski and Einstein, is that the Lorentz transformations are 
written as if they referred to finite magnitudes, when in fact 
they should refer to infinitesimals. The present author believes that to this error can be traced 
most of the so-called ``paradoxes'' that pervade Special Relativity.\\
\indent The Lorentz transformations have the purpose of finding the coordinates 
of an event, from the point of view of one coordinate system, 
given the coordinates of the same event as seen from a second 
coordinate system.\\
\indent Einstein wrote \cite{einstein1916}: ``Any such event 
is represented with respect to the co-ordinate system K by the 
abscissa x and the time t, and with respect to the system K' 
by the abscissa x' and the time t'. We require to find x' and 
t' when x and t are given.'' \\
\indent Einstein wrote the Lorentz transformation in finite form, in Appendix I of his book \cite{einstein1916} and also in his original paper of 1905 \cite{einstein1905} as follows: \\

\begin{equation}
\emph{t'}  = \gamma (\emph{t - V x/c}^{\mathit{2}})
\end{equation}
\begin{equation}
\emph{x'} = \gamma (\emph{x - V t})
\end{equation}
where \emph{V} represents the relative velocity between the two 
frames of reference, let the two frames be called S' and S (Einstein 
named K and K', but that is insignificant). Let us assume that 
the frame S' moves to the right in the \emph{x} direction (or the
frame S moves to the left).

\begin{figure}[htbp]
\begin{center}
\includegraphics[viewport=-50 0 600 320,width=15cm,clip]
{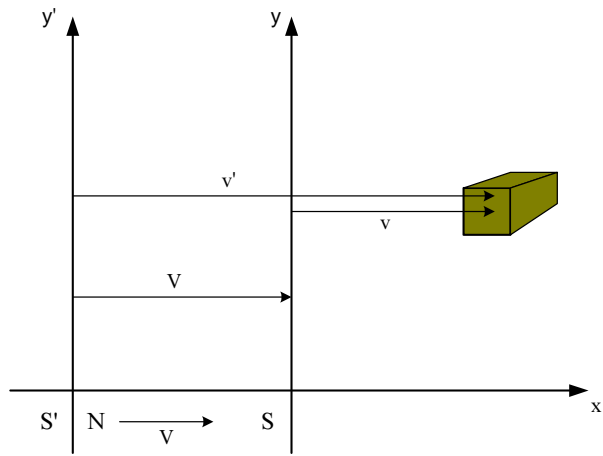} 
\caption{Coordinates}
\end{center}
\end{figure}

The inverses of these equations are:

\begin{equation}
\emph{t}  = \gamma (\emph{t'} + \emph{V x'/c}^{\mathit{2}}) 
\end{equation}
\begin{equation}
\emph{x} = \gamma (\emph{x'} + \emph{V t'})
\end{equation}

\indent These equations can be used, in theory, to compute the coordinates \emph{t'} 
and \emph{x'} of a certain \emph{event}, as seen from the frame of reference 
S', if we know the coordinates \emph{t} and \emph{x} of the same event 
as seen from the frame S, or vice versa. 

Let us try now to reveal that they cannot be used to accomplish this goal.

First, assume that the relative velocity \emph{V} between the two 
frames is zero. Then, the \emph{Lorentz contraction factor}

\begin{equation}
\gamma =\frac{1}{\sqrt{1-\frac{V^{2} }{c^{2} } } } 
\end{equation}\\

becomes equal to 1 and the transformations are simplified 
to \emph{t'} = \emph{t} and \emph{x'} = \emph{x}. This means that the coordinates of the \emph{event} are the same when they are seen from the two frames of reference and when those frames of reference do not have a relative movement between them. It can also be concluded that the two origins are coincident.\\
\indent Now let us assume that the origin 
of S' is displaced a certain given distance 
to the left of the origin of S, as in the previous figure. No matter what the relative velocity between the two frames of reference, but in particular if it is zero, it would be necessary to, somehow, add that distance to the coordinate \emph{x} in order to find \emph{x'} (or subtract that distance from \emph{x'} in order to find \emph{x}).
 
The coordinate transformations that include the mentioned constant distance are called Poincar\'{e} transformations. But, in the same sense, are mistakenly interpreted as that they transform \emph{coordinates} instead of \emph{intervals}. \\
\indent Let us provide some numbers in a given example and then let us do the computations suggested by equations (1) and (2). Assume that the origin of the frame S is at rest in our laboratory; also, the velocity \emph{V} of the frame S' (moving to the right) is so small that, in practice, can be ignored (assume, for example, one meter per day). Now, the problem is to compute the coordinate \emph{x'} at the instant in which the origin of the frame S' is at 100 meters to the left of the origin of the frame of reference S and the coordinate \emph{x} of the \emph{event} (for example a tennis ball hits the floor), is produced at 1000 meters to the right of the origin of the frame S. \\
\indent A quick ``mental'' estimation of the value for \emph{x}', assuming that both frames of reference are static, produces 1100 meters; on the other hand, according to equation (2), the \emph{coordinate x}' is given by: $\gamma$ $(1000$ meters $- 1$ meter/day * \emph{t}), where $\gamma$ is very close to 1 and you should tell me now what is the time \emph{t}. Did you obtain minus one hundred days? Fine.\\
The computation is something like:\\
\begin{equation}
1100m=1000m - 1m/day \cdot (-100day)
\nonumber
\end{equation}
What is the meaning of this time? It is, evidently, the interval of time since the event was produced until the origin of the primed frame of reference reaches the origin of the unprimed frame of reference. From this computation we can conclude that the magnitude of the time variable is greater when the speed is smaller, so as to be close to infinite when the relative velocity of both frames of reference is close to zero. So, one conclusion almost obvious here, is that it is not an independent \emph{time coordinate}, because it must depend on the relative positions of the origins of the two frames of reference at the time when the event occurs and, also, such time depends on the relative velocity between both frames of reference. In this context we realize that the phenomena is behaving, and should be explained, as a differential equation. The supposed \emph{time coordinate} is not such thing.\\
\indent It seems that we still have to do another ``Gedanken Experiment'' to realize what is happening. Let us assume again that the two frames have a relative velocity \emph{V}. 
The argument is simplified a little if we assume a small velocity, 
for example one meter per day. For such a small velocity we 
can ignore $\gamma$ again. But, strictly speaking, it is not necessary 
this simplification because the problem to be revealed here is 
too gross to be overlooked, whatever the value of $\gamma$. Also, 
assume that the event occurs close to you, at the origin of 
the \emph{x}-axis of the S frame. This means that the \emph{x} 
coordinate of the event is \emph{x}=0. With this assumption, the transformation 
(2) simplifies to: $x' = - V t$. Finally, in order to 
define the time \emph{coordinate} of the event, assume that the \emph{event} occurs at the precise instant at which you look at your wristwatch. Thus, the time coordinate, \emph{t}, is precisely the hour you have at this moment in your watch. If, sometime later, one of your neighbors passes near you in his/her car, whose frame of reference is S', and you inform him/her, written in a piece of paper, the coordinates you just recorded for the event in your frame S, he/she could, in theory, compute the coordinate \emph{x'} where 
the event happened, with respect to his/her frame of reference, 
if he/she replaces those coordinates in the original equations (1) and (2), or in the simplified transformation \emph{x'} 
= \emph{- V t}.\\
\indent I am almost sure that he/she will not be able to compute a reasonable result, 
in the first place because the origin of the \emph{time coordinate} is usually different for different observers. For example, the 
Gregorian calendar marks currently a few more than 2000 years, 
a few more than 1380 years in the Islamic chronology and a few 
more than 5760 years in the Jewish calendar. Those numbers make 
nonsense if you try to use them as origins of the temporal \emph{coordinates} in the 
above equations. So, for example, if I have the time coordinate of 2005 years, he/she could have 5765 or 5766 years. It does not 
help much if the beginning of the time coordinate is assumed 
midnight, because it is highly probable that I do not have the 
same hour as you do, due to our respective geographical position. 
And you should not suppose that synchronizing our clocks then you can use the above ``coordinate transformations'' because, even if we synchronize our time \emph{coordinate}, in years, days, hours, and seconds, the product of a \emph{time coordinate} by a velocity, produces nonsense for the problem at hand. \\
\indent The only reasonable way to compute the \emph{interval x'} is if I further specify that the clock is not a clock but a chronometer or stop-watch that can be put to zero when the event takes place. In other words, you need to know what is the \emph{elapsed} time since the occurrence of the event. You can obtain a reasonable result only if you know, for example, that the event took place one hundred days before the origins of the \emph{x} coordinates of both frames became coincident. \\
The conclusion of the above argument is that the variables used 
in equations (1) and (2) are not \emph{coordinates} but \emph{intervals}. If you use space and time \emph{intervals}, which are still finite, those equations work better but not quite. \\
\indent The Lorentz transformations should be written with differentials, because, in that case, the needed constant could enter as a constant of integration. The same can be sustained for the time transformation (1).\\
\indent The mathematicians should explain if it is correct, as is usual and accepted in current Physics, to interpret the Lorentz transformations in differential form as if they were equivalent to the finite transformations, or if a new proof is needed.\\
\indent The Lorentz transformations with the use of differentials would be the following:
\begin{equation}
\emph{dt'}  = \gamma (\emph{dt - V dx/c}^{\mathit{2}})
\end{equation}
\begin{equation}
\emph{dx'} = \gamma (\emph{dx - V dt})
\end{equation}
\indent 
Take note that the \emph{event} is instantaneous and consequently does not have a velocity of displacement associated with it. The variable \emph{V} was used in this section, and should be interpreted, as the relative velocity between the two frames of reference.\\

Please verify that these equations are identical to the equations (1) and (2), except that we are now using differentials instead of intervals.

Let us note that if we look at the \emph{event} from the origin of a third frame of reference, such as S'', then the velocity, \emph{v}'', to be used both explicitly in the equations as well as in the Lorentz contraction, should be now the relative velocity between the frames of reference S'' and S', whose corresponding time and space intervals we are trying to compute now.

If we require to compute both the time and space differentials corresponding to the frame of reference S', given the coordinate differentials defined for the frame of reference S'', we will still have to use equations (6) and (7), with the necessary corrections, which produce the following end result:
\begin{equation}
\emph{dt}'  = \gamma'' (\emph{dt'' - v'' dx''/c}^{\mathit{2}}) \\
\end{equation}
\begin{equation}
\emph{dx}' = \gamma'' (\emph{dx'' - v'' dt''})
\end{equation}
where \emph{v}'' is interpreted currently as the relative velocity of the frame of reference S'' with respect to the frame of reference S' (or vice versa).

Equating equations (6), (8) and (7), (9) we come up with:
\begin{equation}
\emph{dt}'  = \gamma (\emph{dt - V dx/c}^{\mathit{2}}) = \gamma \,'' (\emph{dt'' -- 
v'' dx''/c}^{\mathit{2}}) = \dots
\end{equation}
\begin{equation}
\emph{dx}'  = \gamma (\emph{dx - V dt}) = \gamma \,'' (\emph{dx'' -- v'' dt''}) =  \dots 
\end{equation}

The velocities \emph{V}, \emph{v''}, \dots  associated with each frame of reference are equal to the velocities between the frame of reference S' and each one of the frames of reference S and S'' of the observers (or vice versa). 

The equations for the different $\gamma$'s are functions of the corresponding velocities. The square of a differential of interval, $ds^2$, can be obtained from here by squaring any of the corresponding couple of equalities, and subtracting (11) from (10) multiplied by the square of the speed of light.

Solving the last two equalities (10) and (11) for \emph{dt}'' and \emph{dx}'' we obtain:
\begin{equation}
\emph{dt}\,'' = \Gamma (\emph{dt - v dx/c}^{\mathit{2}}) 
\end{equation}
\begin{equation}
\emph{dx}\, '' = \Gamma (\emph{dx - v dt})
\end{equation}

where \emph{v} is an abbreviation for the expression:
\begin{equation}
\emph{v} = \frac{V-v''}{1-\frac{V \cdot v''}{c^2}}
\end{equation}
which represents the velocity between the frames of reference S and S'' and is comparable with Einstein's equation for ``composition of velocities''.

Whereas $\Gamma$ is defined as:
\begin{equation}
\Gamma = 1/\sqrt{1-\frac{v^2}{c^2}}
\end{equation}

The previous equations reveal the group character of the Lorentz transformations.\\
Equations (12) and (13) can be rewritten and simplified by replacing \emph{dx/dt} by \emph{c}:
\begin{equation}
\frac{dt\,''}{dt}=\frac{dx\,''}{dx}= \sqrt{\frac{1-v/c}{{1+v}/c}}
\end{equation}
The proportion of time differentials has the same expression as the space differentials and represents the Doppler redshift of an event as observed from two arbitrary frames of reference.\\

\section{A simple imagined experiment}

The intent of this section is to give a clear account of the inner 
workings of the Lorentz transformations and to suggest a 
possibly correct interpretation for some of its variables. \\
\indent
A simple imagined experiment is analyzed that gives some insight 
about the need for light to travel at the same speed, according 
to two inertial observers. The interpretation suggested by the author is that the Lorentz transformations predict, and 
need, an invariant speed of light in order to compute correctly 
the space and time intervals corresponding to a given event, 
even though those space and time intervals are different from 
the point of view of each one of the observers.

I have devised the following imagined experiment to illustrate 
how SRT should be applied, and to show that, in such setting, 
it is possible for light to travel at the speed \textit{c} relative 
to two arbitrarily moving observers, in apparent consistency 
with SRT.

The setting is as in the following rough schema:\\

\tab \tab O ------------------------ O'------------{\textbar} E ------------\texttt{>}x\\

Let us assume that there are two coordinate systems with respective 
origins and inertial observers at O and O'. The origins and the 
observers of the two systems are separating with the fraction $V=\sqrt{3}/2$ of the speed of light, \textit{c}, (assume that O' is displacing in the positive \textit{x} direction or O is displacing in the negative \textit{x} direction). Precisely which one of the two systems is moving is not specified because, according to SRT, this is indifferent.\\

The Lorentz contraction factor has the well-known expression 
given by equation (5).

The numerical value for the Lorentz contraction factor is 2, given the above assumed value for the speed \textit{V}.\\

Also, assume that there is some point ``E'' 
that is always fixed at exactly (dx=1*c*sec) meters to the right 
of the origin O (this is dx=299792458 m or roughly 300000 km). 
The point O' is approaching to E at the same speed with which 
the two observers are separating, because the point E is assumed 
rigidly connected to the origin O. 

Finally assume that, at the instant when the origins of the two 
systems are coincident, a photon is emitted from the point of 
the origins towards the point E. The question is: what the time 
and space intervals for the event ``the photon reaches 
the point E'' are? according to both observers.

Let us do now the math. \\
According to the observer at O, it takes for the photon precisely 
one second to reach the point E (dt=1 sec). In this time interval, 
according to whatever observer, the origins of the two systems 
have separated, at the speed V, roughly 260000 km (I am not sure whether SRT is of some use to predict this). 

The equations that SRT should use to compute the time and 
space intervals of the event are (6) and (7). 

The primed values, computed using these equations, are: 
dt'=0.268 sec and dx'=0.268*dx.

In summary, some of the answers provided by SRT are the following:

1)\tab 
The observer at O has remained at rest with respect to the point 
E after the emission of the photon and, consequently, obtains 
the time interval of \textit{1 sec} and the space interval of \textit{1*dx}, 
using the speed of light \textit{c}.

2)\tab 
For the observer at O' we compute that he is moving with respect 
to the points O and E at the speed \textit{V}, and obtain that 
the photon will take the time dt'=0.268 sec to reach the coordinate 
of the point E, which, according to the second equation, is 
at dx'=0.268*dx. Dividing the space interval dx' by the 
time interval dt' we obtain again the same speed c for 
light.\\

3)\tab 
For the observer O assumed at rest, the space interval remaining 
between the points O' and E, after the one second interval, appears 
to be at (1-0.866)*dx=0.133*dx, and the time interval needed 
to cover such distance, at the speed of light, is 0.133 sec.

4)\tab 
For the moving observer, the same space and time intervals mentioned 
in the previous point were computed in point 2 above and 
are double (in fact $\gamma$) the values computed by the observer 
at rest.\\

Take notice that in this SRT experiment we have considered only one-way 
speeds of light. The coordinate intervals of the event E as seen 
from both observers O and O' are computed in an apparently correct 
form if both of them assume the same speed of light. I 
have not developed the computations, if they are possible, assuming 
that O' is at rest and O is moving but the above computations 
seem to be enough to prove that light needs to move with the 
same speed, from the point of view of different moving observers, 
in order to satisfy the postulates of SRT.\\

It is easy to realize that the speeds obtained, in the previous 
experiment, by dividing \textit{dx} by \textit{dt} and \textit{dx}' by \textit{dt}' 
are the speed of light in both cases:

\begin{equation}
\frac{dx'}{dt'} =c, \tab \tab \frac{dx}{dt} =c
\end{equation}\\
This assumption was used at the end of section 1 in order to obtain equation (16).

\section{A new expression for Lorentz contraction factor}

In section 1 of the present paper the author suggested 
that one of the most frequent errors, displayed in the books and 
papers dealing with topics related with relativity theory, is 
to consider the Lorentz transformations as referring to \textit{finite 
coordinates} instead of \textit{interval differentials}. Here the author 
reveals what he considers as the next most frequent errors. 

It is shown that the Lorentz contraction factor, usually represented 
with the Greek letter $\gamma$, does not have the form currently 
accepted for it if such factor is assumed to be equal to the 
quotient of time or space differentials. This error appears very 
frequently in Physics papers and books. See, for example: \cite{debothezat}, \cite{ceapa}, \cite{gift}, \cite{hatch}, \cite{jarrell}, \cite{rutherford}, \cite{schiller}, \cite{selleri}, \cite{smoot}, \cite{takeuchi}, \cite{thide}, \cite{vukelja2}. Of course its origin can be traced back to some Einstein 
writings; see, for example, Appendix 1 from his book ``Relativity: 
The Special And General Theories'' \cite{einstein1916} and the equation 
at the end of section 4 of his original paper about ``The 
electrodynamics of moving bodies'' \cite{einstein1905}. Also, the so-called Lorentz-Fitzgerald ``length contraction'', in its classical form, seems to be another representation of the same error \cite{masud}, \cite{natarajan}, \cite{watson2}.\\

The present author considers that a frequent error 
in relativity theory is to assume that the quotient, between 
space and time differentials, is equal to the relative speed \textit{V} between the two frames of reference, instead of considering such quotient as equal to the speed of light, as shown in equations (17) (from now on let us use \textit{v} instead of \textit{V} for this variable).\\

If the space coordinate, \textit{x}, were given as the speed between the frames of reference, \textit{v}, multiplied by time, as supposed in section 4 of Einstein's paper \cite{einstein1905}, then the proportion between time coordinates does not have the correct form, as the classical Doppler formula (equation (16)), but instead gets the form of the Lorentz contraction factor, as Einstein obtains following such assumption: 

\begin{equation}
\frac{dt}{dt'} =\gamma 
\end{equation}\\
This is different from equation (16). To confirm that this is an error, let us investigate what happen if we divide the Lorentz transformations, equations (6) 
and (7), by \textit{dt} and then divide the first by the second equation:

\begin{equation}
\frac{dx'}{dt'} =\frac{\frac{dx}{dt} -v}{1-\frac{v}{c^{2} } \frac{dx}{dt}
} 
\end{equation}\\

but, according to (17), the space differentials divided by 
the time differentials are equal to the speed of light, so

\begin{equation}
c=\frac{c-v}{1-\frac{v}{c}}
\end{equation}\\

Simplifying, we obtain that the speed of light is equal\dots to  
the speed of light. This is a simple tautology that proves, in 
a partial way, that the differential Lorentz transformations 
are correct. Such proof is partial because the so-called ``Lorentz 
contraction'' factor, $\gamma$, is simplified from both transformation equations when we divide 
one by the other, no matter what the value of this factor.

Let us now treat each of the Lorentz transformations individually.

First, dividing both sides of equation (7) by \textit{dt}, and 
multiplying and dividing by \textit{dt}' in the left hand side, we 
get:

\begin{equation}
\frac{dx'}{dt'} \frac{dt'}{dt} =\gamma \cdot (\frac{dx}{dt} -v)
\end{equation}\\

by replacing equations (17):

\begin{equation}
c\frac{dt'}{dt} =\gamma \cdot (c-v)
\end{equation}\\

Now, let us assume that the ``Lorentz contraction'' 
factor is equal to the quotient of \textit{dt} divided by \textit{dt}', as in equation (18). 
From the following result, which is contradictory with relativity 
theory, it should be obvious that this assumption is an error.

Replacing the quotient of differentials \textit{dt}'/\textit{dt} by the inverse 
of gamma, and solving for gamma, we obtain the following expression:

\begin{equation}
\gamma =\frac{1}{\sqrt{1-\frac{v}{c} } } 
\end{equation}\\

This is a surprising value, different from the one predicted 
in current SRT (see equation (5)). Special relativity theory predicts that the Lorentz 
contraction factor must have the square of the relative speed 
between frames of reference, \textit{v}, divided by the square of 
the speed of light, whereas in this equation we have obtained 
essentially the same result but without the squares.

This result is confirmed using the Lorentz transformation for 
time. Dividing both sides of the equation (6) by \textit{dt} and using equations (17):

\begin{equation}
\frac{dt'}{dt} =\gamma \cdot (1-\frac{v}{c} )
\end{equation}\\

and, replacing equation (23):

\begin{equation}
\frac{dt'}{dt} =\frac{1}{\sqrt{1-\frac{v}{c} } } \cdot (1-\frac{v}{c} )
\end{equation}\\

From equation (25) we confirm that the ratio of time differentials:

\begin{equation}
\frac{dt}{dt'} =\frac{1}{\sqrt{1-\frac{v}{c} } } 
\end{equation}\\

does in fact produce the new form of the ``Lorentz contraction'' 
factor, equation (23). \\
This result is, obviously, inconsistent with the prescriptions 
of SRT.

To obtain the correct form of the ``Lorentz contraction'' 
factor it is necessary that the quotient of time differentials 
be given by the Doppler's formula:

\begin{equation}
\frac{dt}{dt'} =\sqrt{\frac{1+\frac{v}{c} }{1-\frac{v}{c} } } 
\end{equation}\\

If this expression is replaced in equations (22) or (24) it is 
easy to verify that, solving for the Lorentz contraction factor, 
it is obtained the expression predicted by SRT and shown in equation 
(5).

\section{Conclusions}

\tab 
1.\tab I felt that I could be offending the intelligence of the reader by explaining in so much detail elementary examples such as the one analyzed in the section ``the infinitesimal
character of the Lorentz transformation'' that, in principle, can be grasped by any motivated layman, but the case is that, apparently, it has not been recognized, throughout the last one hundred years, that the finite Lorentz transformation should work with intervals and not with coordinates.
Or is the case that physicists, and in particular the ``relativists'' didn't want to recognize it? My suspicion is that we, as humans, have great difficulty in capturing concepts, doing
deductions and considering the cumulus of objections which have been sustained through the years by many investigators. As Feyerabend says: ``the lasting success of our categories and the omnipresence of some specific point of view is not signal of excellence or indicative that truth has been found. Rather \emph{it is indication of the failure of reason} to find adequate alternatives to overcome an intermediate stage of our knowledge'' \cite{feyerabend} (my own translation from the Spanish version of this book). Even though some physicists have been able to identify the many contradictions related with Special
Relativity (\emph{one} should have been enough), they have been incapable of constructing an integrated and coherent theory. Our more advanced reasoning seem to use deductive chains of about five resolution steps. Chains of reasoning with about ten resolution steps are almost impossible for us to find and accomplish without error. An alternative line of investigation, worth of further study, is about the complexity of the knowledge; this should help to pinpoint, and be careful with, arguments where the validation of the truth of the premises and of the chains of deduction are approaching the limits of human comprehension.\\

2.\tab A theory should always be subject to analysis and
improvements. In particular, Special Relativity, as a theory, seems that has not been honest in the estimation of the degree of uncertainty that it was conveying, even though the technical aspects had been better founded. This paper has revealed some \emph{misinterpretations}, which seem to be better qualified as \emph{errors}, that cannot be corrected or improved with experiments but by changing the theory.\\

3.\tab
There are several reasons why SRT has been suspicious during 
all its existence. 

First, SRT is incompatible with the ontology of many philosophers, 
physicists, mathematicians and free thinkers due to the seemingly 
unnatural behavior of some of the basic physical concepts such 
as mass, length and time, that this theory tries to reveal. Some 
of its consequences are correct, although counter intuitive, in the sense that they seem to predict the real outcome of Nature, despite their weird behavior.

\indent Nevertheless, there are some hints to conclude that some of the relativity equations have had sloppy and flawed derivations. Some of such ``derivations'' should not be considered as such, precisely due to the mathematical, 
logical and other kinds of errors made by some relativists and 
their followers. Check, for example, the paper of Aleksandar Vukelja \cite{vukelja} that examines 
the ``Simple Derivation of the Lorentz Transformation'' 
developed by Albert Einstein as a supplementary to section XI 
of his book of 1920 entitled ``The Special and General 
Theory'' \cite{einstein1916}. Vukelja concludes: ``It takes exceptionally 
strong illusions and lack of math skills to make five such errors 
on a single sheet of paper.''\\

4.\tab 
The computations, in the simple numerical setting analyzed in section 2, confirm that SRT requires that the speed of light be the same for different moving observers. More precisely, the space intervals of a given event divided by their corresponding time intervals resulted equal to the speed of light.

And vice versa, it was suggested that such quotients between 
space and time intervals, which appear in the Lorentz transformations, should be interpreted as equal to the speed of light, as shown in equations (17), in order to maintain consistency.\\ 

5.\tab
In section 3 of the present paper it was demonstrated that, if the quotient of time differentials for two moving observers is assumed to be equal to the Lorentz contraction factor, then the expression solved for this factor is not the classical one but something different. This internal contradiction within SRT is one of the most frequent errors in papers dealing with relativity. 

It was suggested that, to restore consistency, it is necessary to reject both the Einstein's speed equation and the quotient of time differentials as equal to the $\gamma$ factor. With the name ``speed equation'' is meant the quotient between space and time \textit{coordinates} (according to Einstein), which should not be equal to the speed \textit{v} between frames of reference but equal to the speed of light. \\

6.\tab
There is a huge number of contradictions or ``paradoxes'' 
in relativity that should have been avoided by means of some 
more careful derivations. Even one small improvement could have 
avoided most of the paradoxes. In the present paper it has been shown one of those improvements, where it is evidenced that 
the Lorentz transformations must \textit{not} use \textit{finite coordinates} but \textit{infinitesimal intervals}.

It is a well-known fact that Special Relativity is based on the \emph{finite coordinate} transformations, which can be traced back to Voigt, Lorentz and Minkowski but popularized by, essentially, the original works of Einstein. As these transformations have been here proved wrong, because they should use neither finite nor coordinate distances, many of the applications and paradoxes based in such equations collapse and disappear.

What every scientist should understand, but in this particular case the physicists, is that \textit{one} contradiction is enough to invalidate a whole theory (or at least some of its premises 
or conclusions). What is not easy, and sometimes takes one hundred years, is to spot the premises or the sloppy reasoning that cause the contradictions. \\

7.\tab
After the error is pinpointed it is ``obvious''. 
However, it will take a lot of time, if ever, to convince some people about it and to correct the several thousands of references returned with a simple query to Google, such as ``Lorentz Fitzgerald contraction'' (without the quotes). The first page of answers that I receive to that query shows a few encyclopedias and dictionaries, such as Britannica, Columbia, Infoplease, Wikipedia, that define it as ``The hypothesis held that any material body is contracted in the direction of its motion by a factor $\sqrt{1-v^2/c^2}$, where \textit{v} is the velocity of the body and \textit{c} is the velocity of light.'' If my previous arguments are correct then the said contraction should not have this form but the one of equations (16) or (27).

One must bear in mind that, according to Evert Jan Post \cite{janpost}, ``it requires honesty and courage to reject an ontic invention when it leads to contradictions and has been proven to be unworkable.''
\\

\end{document}